\documentclass{PoS}

\title{On the Lefschetz thimbles structure of the Thirring model}

\ShortTitle{On the Lefschetz thimbles structure of the Thirring model}

\author{\speaker{Kevin Zambello} and Francesco Di Renzo\\
        Dipartimento di Scienze Matematiche, Fisiche e Informatiche, Universit\`a di Parma and INFN, Gruppo Collegato di Parma, I-43124 Parma, Italy\\
        E-mail: \email{kevin.zambello@pr.infn.it}, \email{francesco.direnzo@pr.infn.it}}

\abstract{
The complexification of field variables is an elegant approach to attack the sign problem. In one approach
one integrates on Lefschetz thimbles: over them, the imaginary part of the action stays constant and can be
factored out of the integrals so that on each thimble the sign problem disappears. However, for systems in
which more than one thimble contribute one is faced with the challenging task of collecting contributions
coming from multiple thimbles.
The Thirring model is a nice playground to test multi-thimble integration techniques; even in a low dimensional
theory, the thimble structure can be rich. It has been shown since a few years that collecting the
contribution of the dominant thimble is not enough to capture the full content of the theory. We report
preliminary results on reconstructing the complete results from multiple thimble simulations.}

\FullConference{37th International Symposium on Lattice Field Theory - Lattice2019\\
		16-22 June 2019\\
		Wuhan, China}

\usepackage{placeins}
\begin{document}

\section{Introduction}
The sign problem is a major obstacle hindering the exploration of the QCD phase diagram by lattice simulations. In a lattice simulation,
the task of computing an observable amounts to computing an expectation value $\langle O \rangle = \frac{1}{Z} \int dU O e^{-S}$ with respect
to the Boltzmann weight $e^{-S}$, where $S$ is the action of the system. Since at finite density the QCD action is complex valued, 
one cannot sample configurations $\propto$ $e^{-S}$ to estimate observables by importance sampling.
A huge effort is ongoing to find a way to overcome the sign problem and an elegant approach is the complexification of field variables.
This approach is being explored by various techniques, i.e. the complex Langevin method \cite{sinclair:lattice2019} \cite{tsutsui:lattice2019}, thimble regularisation
\cite{direnzo:lattice2019} \cite{ziesche:lattice2019}, the generalized thimble method \cite{alexandru:lattice2019} \cite{valgushev:lattice2019} \cite{fukuma:lattice2019}
and the path optimization method \cite{ohnishi:lattice2019} \cite{mori:lattice2019}.

In thimble regularisation one replaces the original integration path by a set of manifolds called thimbles. Since over thimbles the imaginary
part of the action stays constant, integrals on each thimble can be efficiently estimated by importance sampling. Nonetheless, one also needs
to know the relative weights of the contributions coming from different thimbles. Calculating these weight has proven to be a tricky point,
though some computation methods have been proposed and applied in simple models \cite{DiRenzo:2017igr} \cite{Zambello:2018ibq} \cite{Bluecher:2018sgj}.

Here we apply thimble regularisation to the $1$-dimensional Thirring model. This theory has a rich thimble structure and it has been shown before
that the contribution coming from the dominant thimble is not enough to reproduce the exact results \cite{Fujii:2015vha} \cite{Alexandru:2015sua}.
Firstly we show that the one-thimble approximation indeed fails at strong couplings. Then we try to collect the contribution coming from the sub-dominant thimble.
Finally we explore the concept of Taylor expansions applied to thimble regularisation.

\section{The Thirring model}
\subsection{The theory}
On the lattice, the $1$-dimensional Thirring model is defined by the action
$$S = \beta \sum_n (1 - cos(x_n)) - log~detD$$
$$detD = \frac{1}{2^{L-1}} \left(cosh(L \hat{\mu} + i \sum_n x_n) + cosh(L~asinh(\hat{m})) \right) \mbox{ . }$$
In the above expressions the sums run over the sites of a one-dimensional lattice of length $L = N_t$, $x_n$ is the discretization of an auxiliary bosonic field,
$\beta = \frac{1}{2 g^2}$ is the (half inverse squared) coupling constant, $\hat{\mu}$ and $\hat{m}$ are respectively the chemical potential and the fermion mass in lattice units.

For this theory the partition function can be solved analytically, hence analytical solutions are also known for the observables number density and chiral condensate,
$$\langle n \rangle = \frac{1}{L} \frac{\partial log~Z}{\partial \hat{\mu}} = \frac{I_1(\beta)^L sinh(L \hat{\mu})}{I_1(\beta)^L cosh(L \hat{\mu}) + I_0(\beta)^L cosh(L~asinh(\hat{m}))}$$
$$\langle \bar{\chi}\chi \rangle = \frac{1}{L} \frac{\partial log~Z}{\partial \hat{m}}= \frac{1}{cosh(asinh(\hat{m}))} \frac{I_0(\beta)^L sinh(L~asinh(\hat{m}))}{I_1(\beta)^L cosh(L \hat{\mu}) + I_0(\beta)^L cosh(L~asinh(\hat{m}))} \mbox{ . }$$

In order to thimble regularise the theory, we need to complexify the degrees of freedom first, $x_n \mapsto z_n = x_n + i y_n$. Secondly, we need to identify the critical
points. This can be accomplished by imposing the stationary condition
$$\frac{\partial S}{\partial z_n} = \beta sin(z_n)  - i \frac{sinh(L \hat{\mu} + i \sum z_n)}{cosh(L \hat{\mu} + i \sum z_n) + cosh(L~asinh(\hat{m}))} = 0 \mbox{ . }$$
One can see that the sine of $z_n$ must take the same value $\forall n$, therefore the critical points are field configurations
having the same value $z$ on every site apart from a number $n_-$ of sites where the field may take the value $\pi - z$.
Critical points can be labeled by a pair $(n_-, z)$: the integer number $n_-$ can take any value from $0$ to $\frac{L}{2}$,
while the complex values allowed for $z$ are obtained by solving numerically the stationary condition equation at fixed $n_-$.

The thimble decomposition for this model was discussed in ref. \cite{Fujii:2015bua}. In presence of a Stokes phenomenon between two critical points, there is a change in the
intersection numbers associated to those critical points. By studying the imaginary part of the action as a function of $\mu$ for the critical points of the theory,
one can look for Stokes phenomena and obtain information on how the thimble decomposition changes.

Following ref. \cite{Fujii:2015bua}, let's consider for instance the case of $L=4$, $m = 1$ and $\beta = 1$.
Fig. $1$ shows how the critical points in the $n_-=0$ sector move as $\mu$ increases (left) and how the imaginary part of the action changes (center).
Only the critical points lying in the left-half complex plane are shown, since because of a symmetry the others give conjugate contributions.
At $\mu = 0$ only the critical points $\sigma_0$ and $\sigma_{\bar{0}}$ have non-zero intersection numbers.
\footnote{Actually this statement is true if we add a small imaginary part to $\beta$, otherwise there is a Stokes between the two critical points and the thimble decomposition is not well-defined.}
At $\mu \approx 0.40$ there is a Stokes between $\sigma_2$ and $\sigma_{\bar{0}}$, the former enters the thimble decomposition and the latter exits the decomposition.
At $\mu \approx 0.56$ there is a Stokes between $\sigma_1$ and $\sigma_0$ and $\sigma_1$ also acquires a non-zero intersection number.
At $\mu \approx 0.73$ there is a Stokes between $\sigma_1$ and $\sigma_2$ and $\sigma_2$ leaves the thimble decomposition.
From fig. $1$ (right) one can also see that, when $\sigma_{\bar{0}}$ and $\sigma_2$ enter the thimble decomposition, the real part of the action on these critical points is
much larger than the real part of the action on $\sigma_0$ and $\sigma_1$, therefore their contribution is exponentially suppressed.

\FloatBarrier
\begin{figure}[tbp]
        \centering
        \includegraphics[scale=0.275]{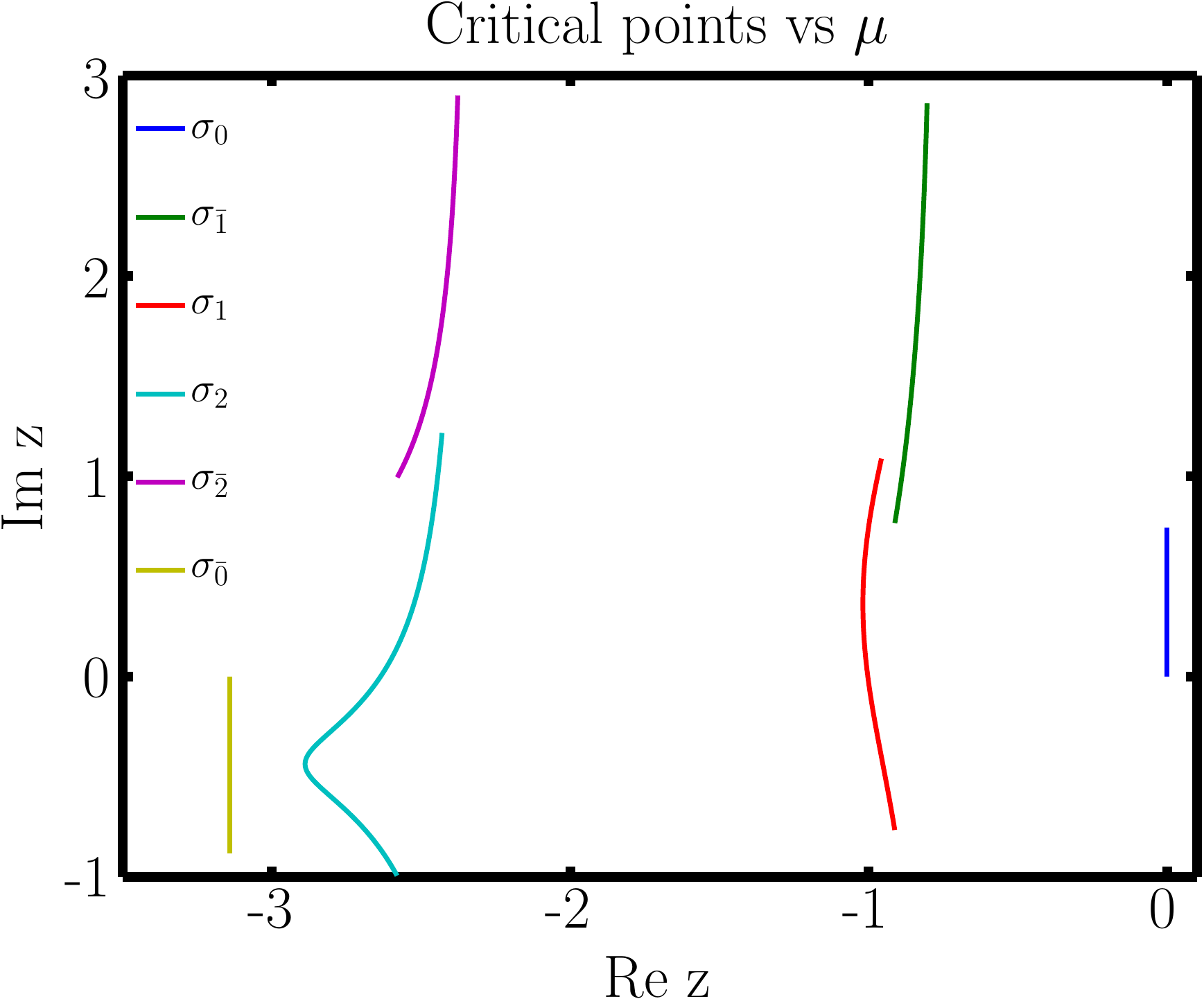}
        \hfill
        \includegraphics[scale=0.275]{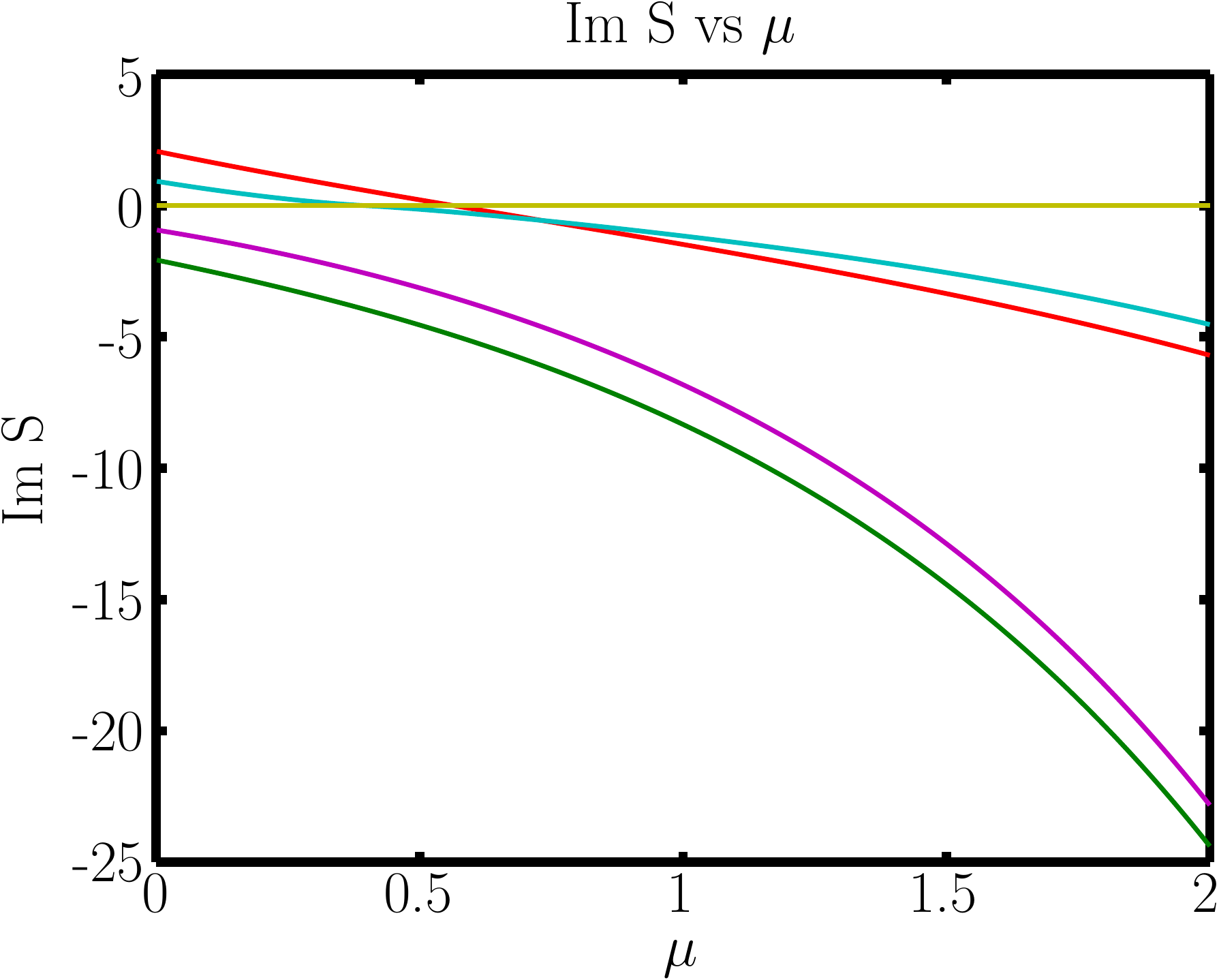}
        \hfill
        \includegraphics[scale=0.275]{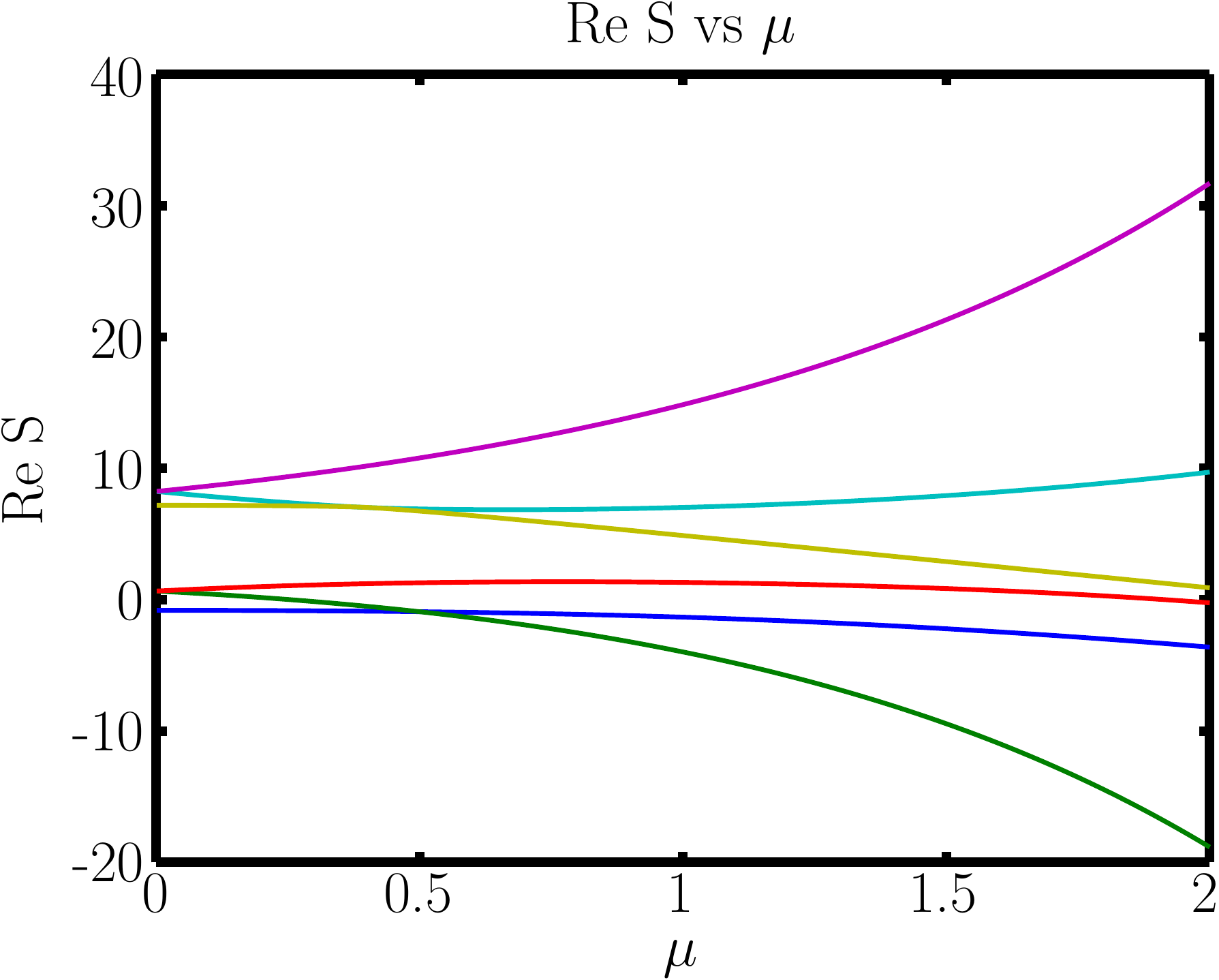}
        \caption{\label{fig:nr1} Critical points in the $n_-=0$ sector for $m=1$, $L=4$ and $\beta=1$. Figures show how the critical points
                                 move as $\mu$ increases (left) and the imaginary part (center) and real part (right) of the action evaluated on those critical points.}
\end{figure}
\FloatBarrier

\subsection{Numerical results from one-thimble simulations}
We ran one-thimble simulations for $L=4$, $m=1$ and $\beta = 1$, $2$, $4$. The chemical potential was varied from $\mu = 0.15$ to $\mu = 1.95$.
Simulations were performed on the dominant thimble, the one attached to the critical point $\sigma_0$.
Fig. $2$ displays the analytical solutions (solid lines) and the numerical results from the simulations (data bars) for two observables,
the number density and the chiral condensate.

The one-thimble approximation works well for weak couplings, but it fails at strong couplings in the transition region.
Discrepancies between the analytical solution and the contribution coming from the dominant thimble are visible
for both observables at $\beta = 2$ and they are especially noticeable at $\beta = 1$.

\FloatBarrier
\begin{figure}[h!]
        \centering
        \includegraphics[scale=0.40]{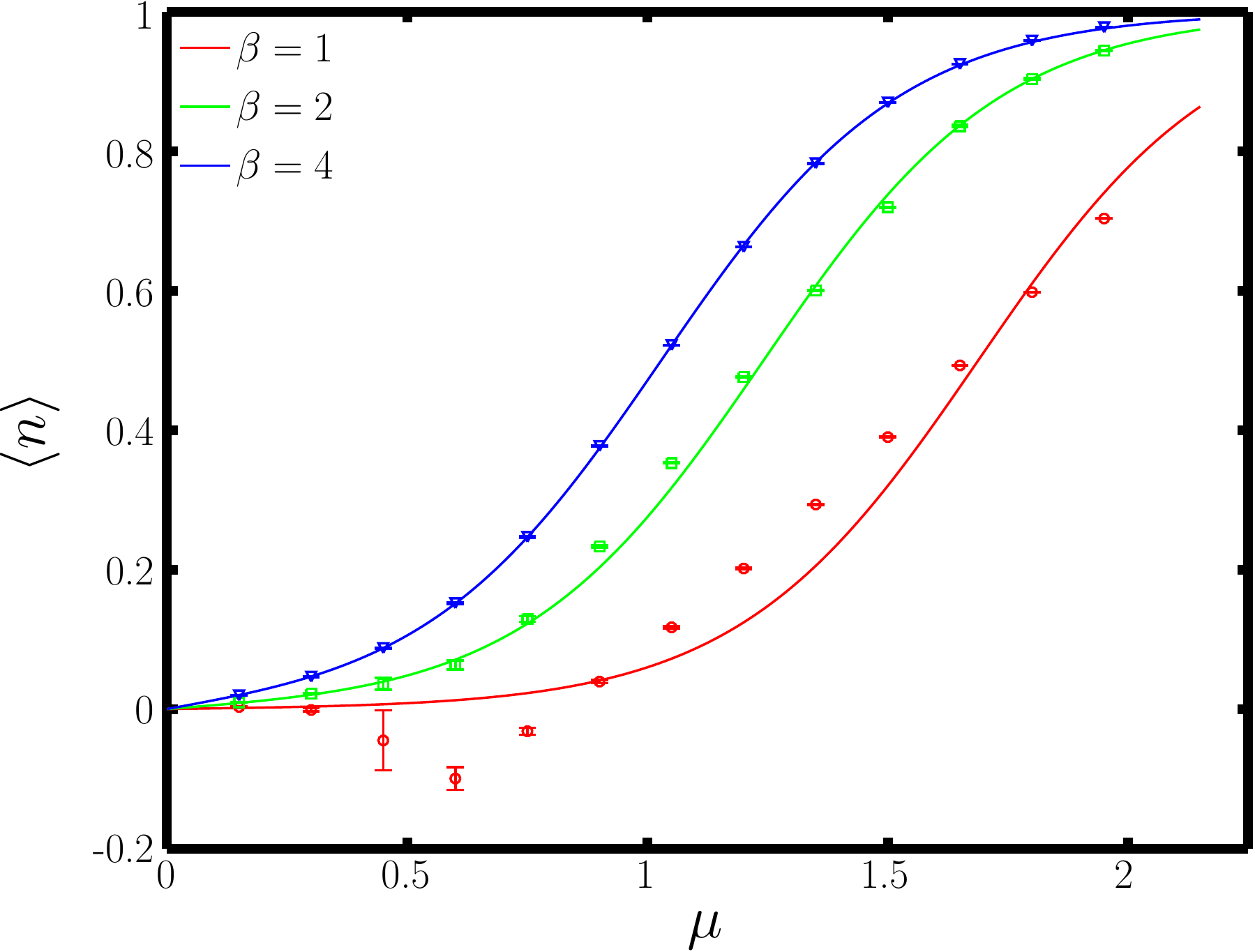}
        \hfill
        \includegraphics[scale=0.40]{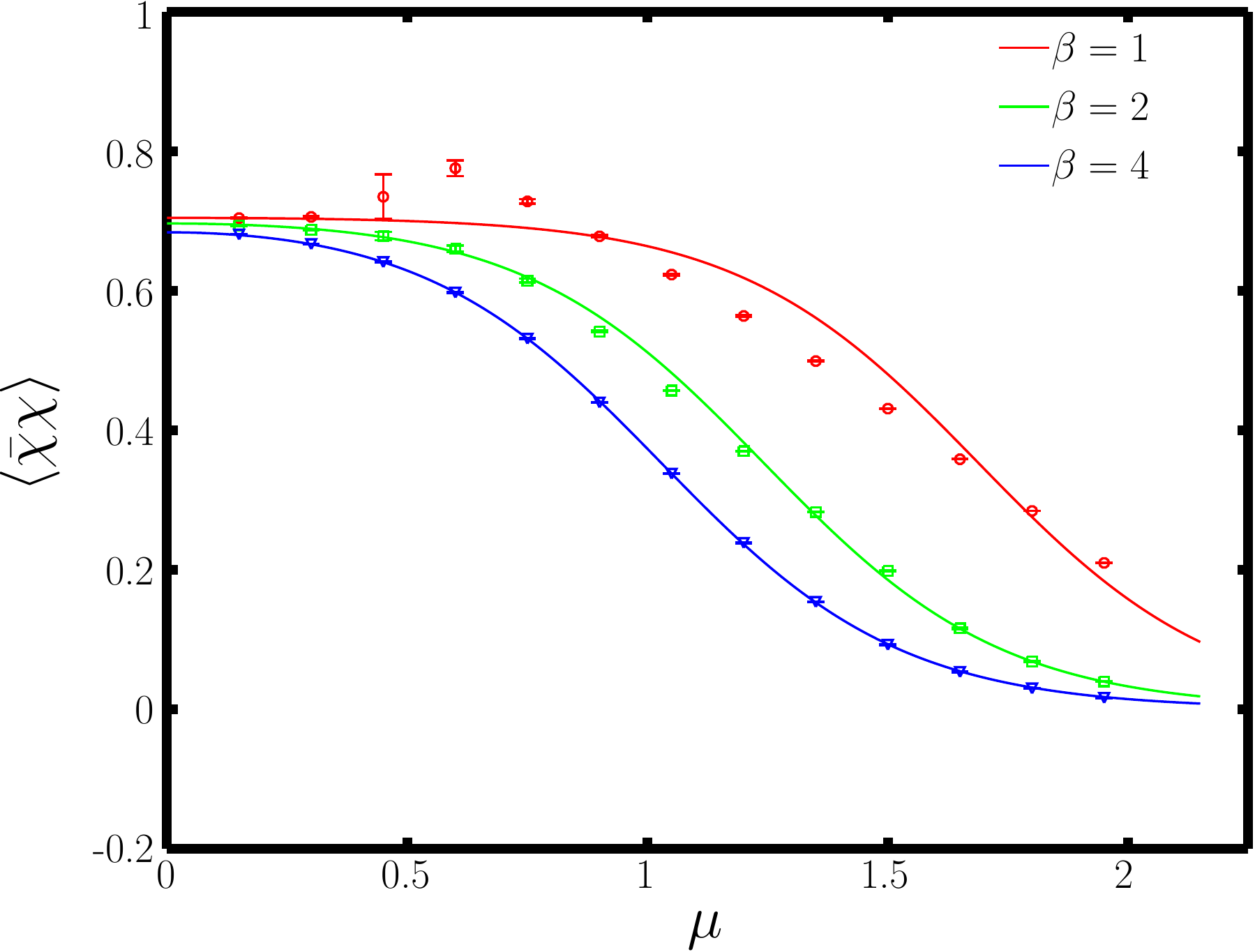}
        \caption{\label{fig:nr2} Results obtained for the number density (left) and the chiral condensate (right) for $m=1$, $L=4$ and $\beta = 1$, $2$, $4$ (red, green, blue).}
\end{figure}
\FloatBarrier

\subsection{Numerical results from two-thimble simulations}
In order to address the discrepancies observed in the transition region at low $\beta$, we tried to collect the contributions coming from both the dominant thimble and
the sub-dominant thimble. The latter is the one attached to the critical point $\sigma_1$.

To obtain the relative weight of the contributions we applied the method
proposed in ref. \cite{DiRenzo:2017igr}: when two thimbles contribute, the multi-thimble decomposition has the form
$$\langle O \rangle = \frac{Z_0 \langle O e^{i\omega} \rangle_0 + Z_1 \langle O e^{i\omega} \rangle_1}{Z_0 \langle e^{i\omega} \rangle_0 + Z_1 \langle e^{i\omega} \rangle_1} = 
\frac{\langle O e^{i\omega} \rangle_0 + \alpha \langle O e^{i\omega} \rangle_1}{\langle e^{i\omega} \rangle_0 + \alpha \langle e^{i\omega} \rangle_1}$$
and the relative weight $\alpha$ of the two contributions can be determined by requiring that the thimble decomposition gives the correct result for an observable already known.
In this case the weight used to compute the chiral condensate was obtained from the analytical solution for the number density and vice versa. In the following
we make use of formalism and notation described in ref. \cite{DiRenzo:2017igr}.

A look at the profile of the partial partition function $Z_{\hat{n}}$ 
\footnote{In our Monte Carlo we sample entire SA paths from the weight $\propto Z_{\hat{n}}$ by importance sampling. The partial partition function $Z_{\hat{n}}$ is
as an integral over the path whose initial direction on the tangent plane is defined by the versor $\hat{n}$.}
for $\beta = 1$ and $L=2$ provided some insights that have been useful to perform the Monte Carlo simulations
for $L=4$.  Fig. $3$ shows the partial partition function for $\sigma_0$ as a function of $n_0$, the component of the initial displacement on the tangent space along the direction associated to
the largest Takagi value. Sharp peaks show up at given values of $n_0$. For $\sigma_1$ the peaks are so thin that relevant configurations are not representable in double precision.
Accordingly, for the Metropolis step we proposed new configurations by choosing smaller rotations between the sub-spaces involving $n_0$ than for the other sub-spaces.
For $\sigma_1$ we also ran the simulations in quadruple precision.

Results are shown in fig. $4$. The region $\mu \sim 0.60 \div 0.75$ is still affected by large error bars for $\beta = 1$, but elsewhere one can see that the discrepancies
due to the the one-thimble approximation effectively disappear after taking into account the sub-dominant thimble.

\FloatBarrier
\begin{figure}[h!]
        \centering
        \includegraphics[scale=0.40]{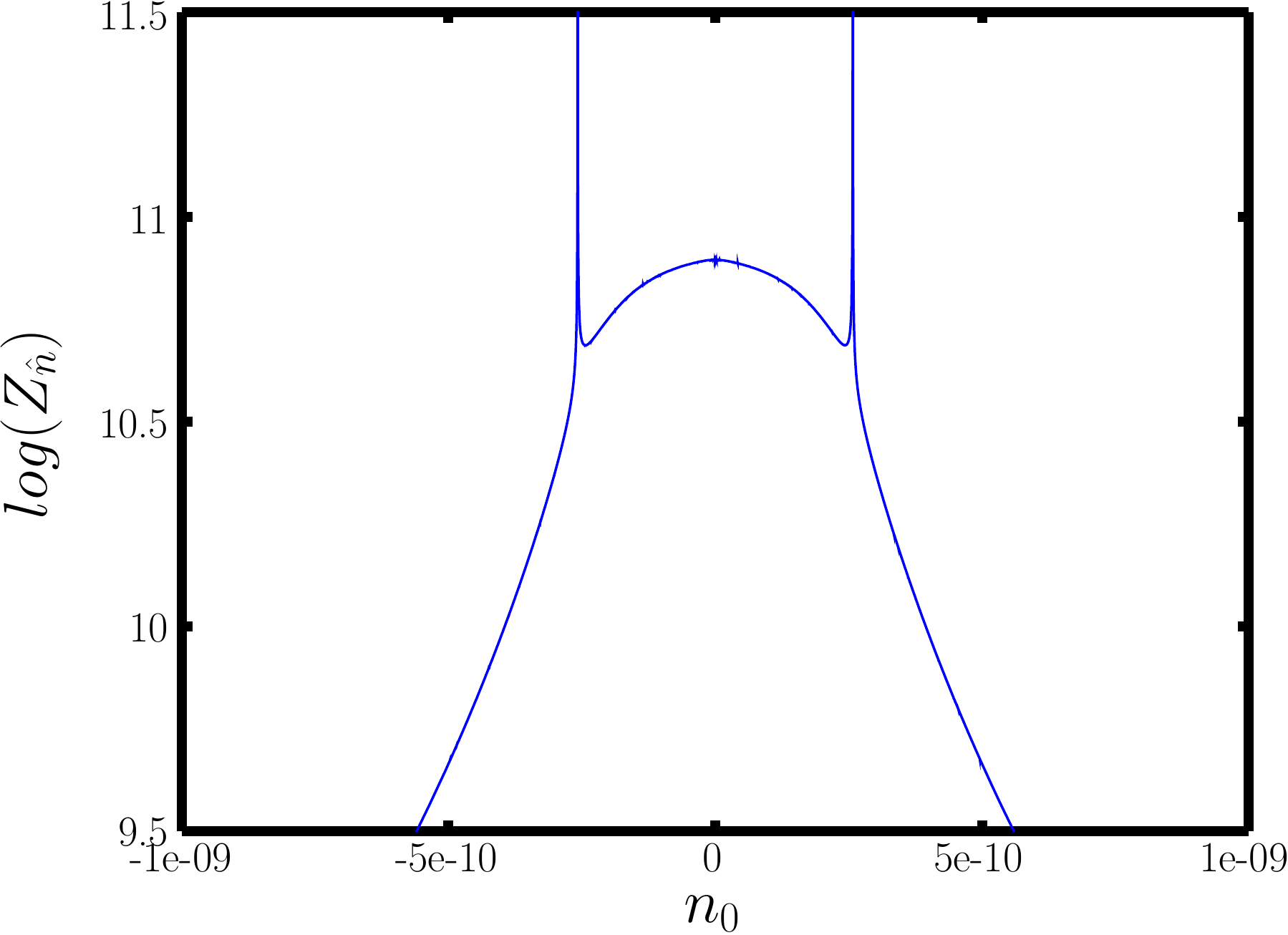}
        \caption{\label{fig:nr3} Profile of $log(Z^{\scriptscriptstyle{(\sigma_0)}}_{\hat{n}})$ as a function of $n_0$ for $m = 1$, $L = 2$, $\beta = 1$ and $\mu = 1.05$.}
\end{figure}
\begin{figure}[h!]
        \centering
        \includegraphics[scale=0.40]{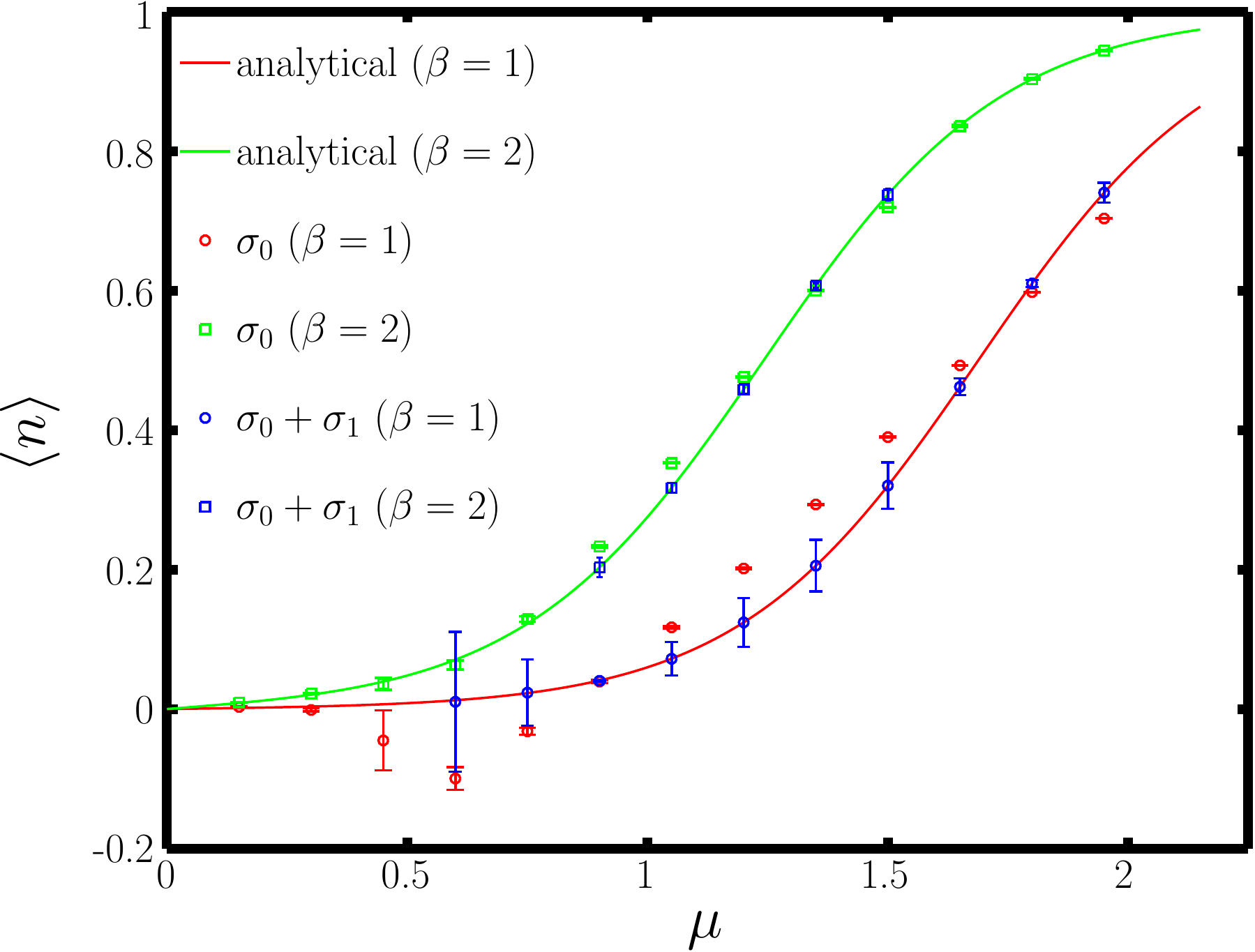}
        \hfill
        \includegraphics[scale=0.40]{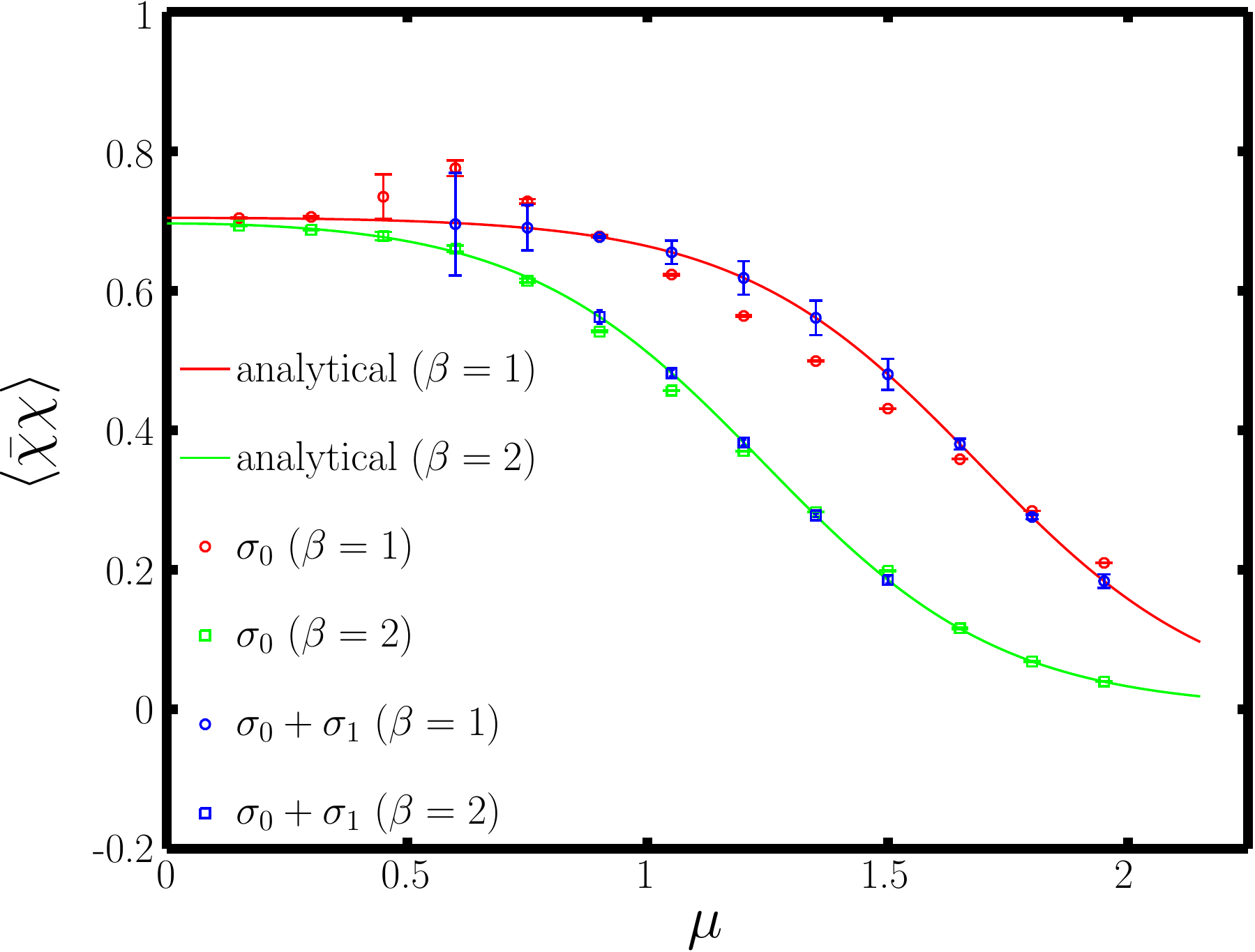}
        \caption{\label{fig:nr4} Results obtained for the number density (left) and the chiral condensate (right) for $m=1$, $L=4$ and $\beta=1, 2$. Results from $1$ thimble are displayed
                                 respectively in red ($\beta = 1$) and green ($\beta = 2$). Results from $2$ thimbles are displayed in blue ($\beta = 1$, $2$).}
\end{figure}
\FloatBarrier

\subsection{The Taylor expansion method}
The Thirring model is an example, among others, of a theory where by
the one-thimble approximation one does not fully recover the exact results.
While this approximation works well at low chemical potentials, in the transition region more than one thimble give relevant contributions inside the multi-thimble decomposition.
In this case, one way to proceed is to try collecting the contributions from more than one thimble, as we've done in the previous section.

However, one could also think of an alternative way (see ref. \cite{direnzo:lattice2019}, an article is also being drafted).
In general observables are continuous, even when the multi-thimble decomposition is discontinuous in presence of Stokes phenomena.
Having that in mind, it is possible to Taylor expand an observable around a point $\mu_0$ where only one thimble gives a relevant contribution:

$$\langle O \rangle (\mu)= \langle O \rangle (\mu_0) + \left. \frac{\partial \langle O \rangle}{\partial \mu} \right|_{\mu_0} (\mu - \mu_0) + \frac{1}{2} \left. \frac{\partial^2 \langle O \rangle}{\partial \mu^2} \right|_{\mu_0} (\mu - 
\mu_0)^2 + \ldots \mbox{ . }$$

The values of the observable at a chemical potential $\mu$ can then be obtained by computing the Taylor coefficients at $\mu_0$ in one-thimble simulations.

We tested such idea on the Thirring model for $m = 1$, $L = 2$ and $\beta = 1$. Results are displayed in fig. $5$. Displayed on the left are the results 
obtained by collecting the contribution from the dominant thimble.
The general picture is similar to what happens for $L=4$: at low and high chemical potentials the only relevant contribution 
comes from the dominant thimble, but this is not true in the transition region. On the right are the results coming from a third order Taylor expansion
around $\mu = 0.15$. After the conference we made further progress by computing a third order Taylor expansion around $\mu = 1.95$, also shown in fig. $5$.
This allowed us to effectively cover the entire range we've explored in $\mu$. The Taylor coefficients have been computed from one-thimble simulations.

\FloatBarrier
\begin{figure}[h!]
        \centering
        \includegraphics[scale=0.35]{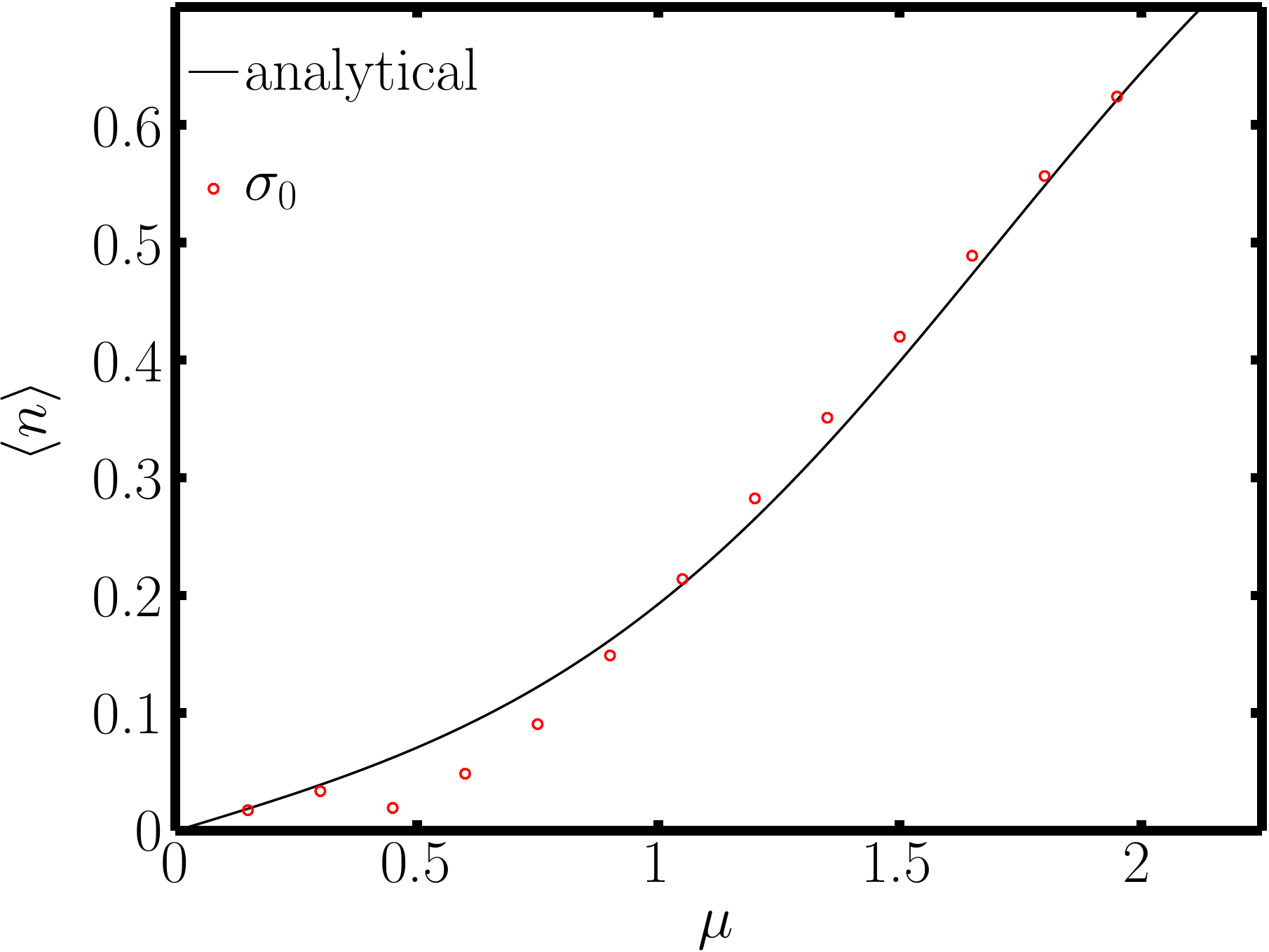}
        \hfill
        \includegraphics[scale=0.35]{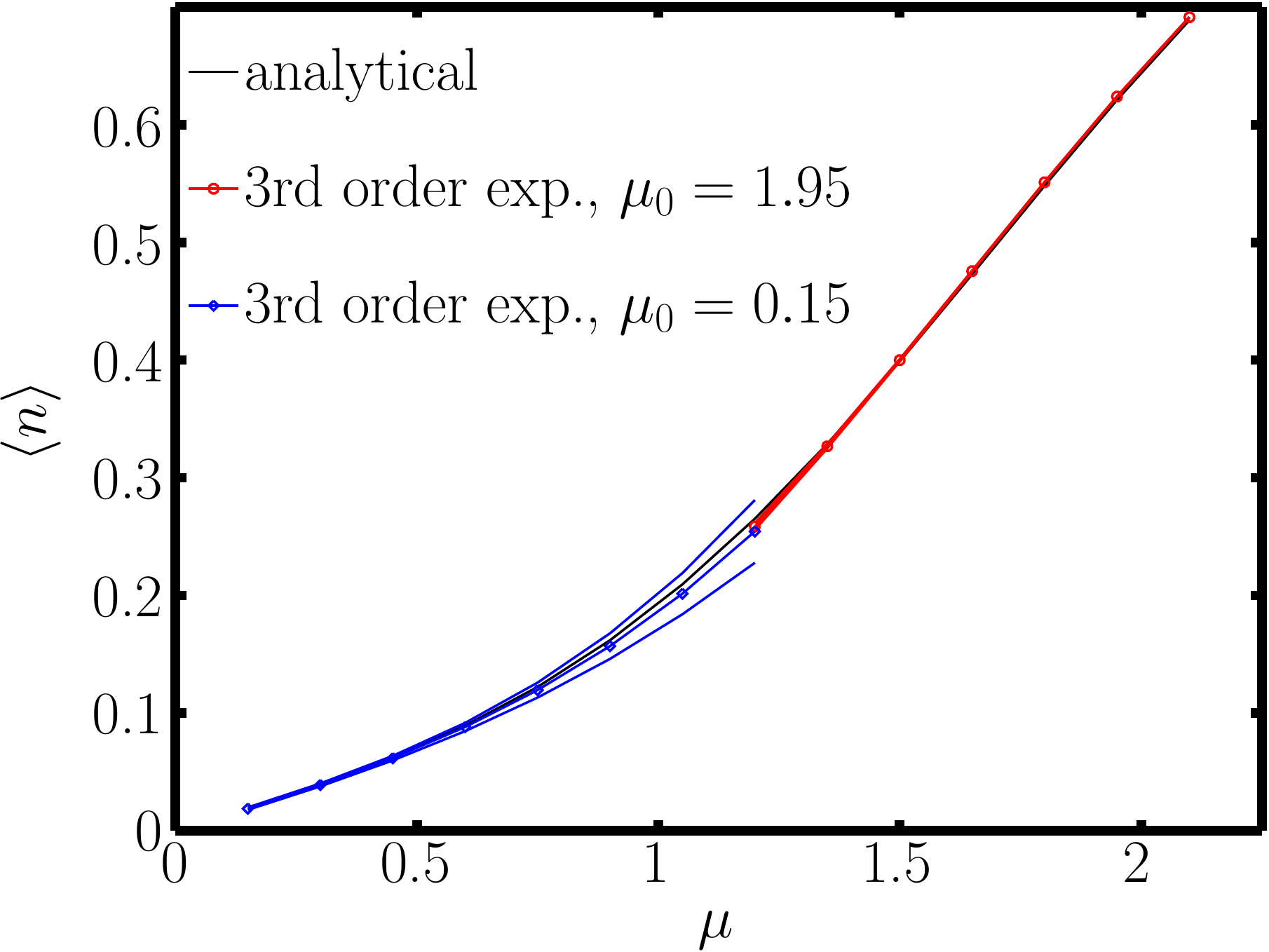}
        \caption{\label{fig:nr5} Results obtained for the number density, $m = 1$, $L = 2$ and $\beta = 1$.}
\end{figure}
\FloatBarrier

\section{Conclusions}
We applied thimble regularisation to the $1$-dimensional Thirring model. By performing one-thimble simulations at $m=1$, $L=4$ and $\beta = 1$, $2$, $4$ we
have confirmed that numerical results do not reproduce the exact ones in the transition region at low $\beta$. We have recovered the exact results
by collecting the contribution from the sub-dominant thimble. We have also explored the concept of Taylor expansion applied to thimble regularisation.
We have tested such idea with parameters $m=1$, $L=2$ and $\beta = 1$ and we have recovered the exact results from one-thimble simulations for chemical
potentials where the one-thimble approximation does not hold.

\section*{Acknowledgments}
This work has received funding from the European Union's Horizon 2020
research and innovation programme under the Marie Sk\l{}odowska-Curie
grant agreement No. 813942 (EuroPLEx). We also acknowledge support from I.N.F.N. 
under the research project {\sl i.s. QCDLAT}.

\end{document}